\documentclass[11pt]{article}

\usepackage[pdftex]{color}
\usepackage[pdftex]{graphicx}
\usepackage{epstopdf,epsfig}
\usepackage{amsmath}
\usepackage{amsxtra}
\usepackage{amssymb}
\usepackage{float}
\usepackage{calc}

\usepackage[mathletters]{ucs}
\usepackage[utf8x]{inputenc}

\definecolor{orange}{rgb}{1,0.5,0}
\definecolor{darkgreen}{rgb}{0,0.5,0}
\definecolor{darkred}{rgb}{0.65,0,0}

\newcommand{\fuder}{\Gamma}
\newcommand{\Det}{\operatorname{det}}

\newcommand{\Grun}{\mathcal{G}}
\newcommand{\datl}[3]{(\frac{\partial #1}{\partial #2})_{#3}}
\newcommand{\dats}[3]{(\partial #1/\partial #2)_{#3}}
\newcommand{\ddatl}[3]{(\frac{\partial^2 #1}{\partial #2^2})_{#3}}

\newcommand{\dhugl}[2]{\datl{#1}{#2}{\mathcal{O}}}
\newcommand{\dhugs}[2]{\dats{#1}{#2}{\mathcal{O}}}

\newcommand{\jump}[1]{[#1]}

\renewcommand{\vec}[1]{\mathbf{#1}}

\newcommand{\bet}{\beta}
\newcommand{\betu}{\bet_0}

\newcommand{\turn}{\theta}

\newcommand{\avg}[1]{\bar{#1}}

\newcommand{\Mach}{M}

\newcommand{\Machu}{\Mach_0}

\newcommand{\uvar}{u}

\newcommand{\uu}{\vec\uvar}
\newcommand{\uuu}{\uu_0}
\newcommand{\ux}{\uvar^x}
\newcommand{\uy}{\uvar^y}
\newcommand{\ut}{\uvar^t}
\newcommand{\utu}{\ut_0}
\newcommand{\utq}{(\ut)^2}
\newcommand{\un}{\uvar^n}
\newcommand{\unq}{(\un)^2}

\newcommand{\unu}{\un_0}
\newcommand{\unuq}{(\unu)^2}

\newcommand{\jn}{j^n}
\newcommand{\jnu}{\jn_0}

\newcommand{\jnq}{(j^n)^2}

\newcommand{\epm}{E}
\newcommand{\epmt}{\hat\epm}
\newcommand{\epmx}{\check\epm}
\newcommand{\spm}{S}
\newcommand{\spms}{s}
\newcommand{\pp}{P}
\newcommand{\ppc}{\pp_c}
\newcommand{\idens}{V}
\newcommand{\idensu}{\idens_0}
\newcommand{\idensc}{\idens_c}
\newcommand{\dens}{\varrho}

\newcommand{\densu}{\dens_0}
\newcommand{\bidens}{b}
\newcommand{\awaals}{a}
\newcommand{\temp}{T}
\newcommand{\rtemp}{t}
\newcommand{\tempc}{T_c}
\newcommand{\rtempc}{t_c}
\newcommand{\tempu}{\temp_0}
\newcommand{\csnd}{c}

\newcommand{\Rspec}{R_s}
\newcommand{\Runiv}{R_u}

\newcommand{\cvspec}{c_v}
\newcommand{\gisen}{γ}

\newcommand{\csep}{\quad,\quad}

\newcommand{\defm}[1]{\emph{#1}}
\newcommand{\subeq}[2]{\mathord{\underbrace{\mathop{#1}}_{#2}}}

\newcommand{\vv}{\uu}

\newcommand{\dnconv}{\searrow}
\newcommand{\conv}{\rightarrow}

\newcounter{eqno}[section]

\newif\ifGPblacktext
\GPblacktextfalse

\title{Van der Waals shock polars with multiple or supersonic critical points}
\author{Volker W. Elling}

\begin{document}

\maketitle

\begin{abstract}
  It is shown that the $\gamma$-van der Waals equation of state (eos) permits shock polars with supersonic critical points,
  corresponding to critical or strong-type shock reflections that are supersonic, which is not possible for ideal gas.
  It is also shown that general van der Waals eos permit polars with multiple critical points,
  corresponding to four or more reflected shocks for same deflection angle.
  Of these reflected shocks at least two are weak-type, i.e.\ deflection angle increasing with increasing shock strength,
  so that standard literature has no criteria to select one of the two.
  Both phenomena can be found with Hugoniot curves entirely in the region of convex and thermodynamically stable eos, avoiding the coexistence region and satisfying various shock stability criteria. 
\end{abstract}

\begin{figure}
  \centerline{\fbox{\footnotesize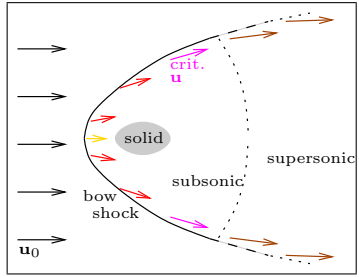}}
  \caption{Bow shock ahead of a blunt body 
    in supersonic flow}
  \label{fig:bowshock}
\end{figure}

Shock polars are fundamental in shock interaction and reflection. 
Consider supersonic flow onto a blunt body (fig.\ \ref{fig:bowshock}). The resulting bow shock ahead of the body has constant upstream state,
but runs through every possible shock angle, from negative to positive Mach angle through vertical (normal shock).
The curve of resulting downstream velocities $\uu$ is called \defm{shock polar} (fig.\ \ref{fig:polypolar}).

\begin{figure}
  \centerline{\footnotesize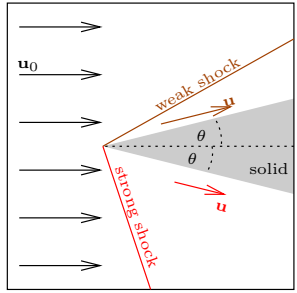}
  \caption{Supersonic flow onto wedge}
  \label{fig:wedge}
\end{figure}

If the body is instead a sufficiently narrow wedge, then straight shocks are attached to the wedge tip (fig.\ \ref{fig:wedge});
they turn upstream velocity $\uuu$ by an angle $\turn$ to make $\uu$ parallel to the wedge surface.
For sufficiently small $\turn$ it is apparent from inspection of the shock polar that on each side there are two possible angles for the tip shock, \defm{weak} and \defm{strong}; 
the weak shock is usually observed. As $\turn$ increases to the \defm{critical angle} the weak and strong shock merge into the \defm{critical shock};
for any larger $\turn$ flows with tip-attached shocks are not possible. 

The shock polar is also essential for many closely related reflection problems, for example two incident into two outgoing shocks, meeting in absence of solid boundaries, 
either in a regular reflection or in a Mach reflection (\cite{neumann-1943,ben-dor-book,hornung-reviews,elling-rrefl-lax});
for these problems the $\pp$-$\turn$ polar is often preferred (fig.\ \ref{fig:superptheta}). 

\begin{figure}
  \centerline{\footnotesize\input{polar.xxx}}
  \caption{Shock polar ($\gisen=7/5$ polytropic, $\Mach_0=2.5$), symmetric across $\uu_0$ axis}
  \label{fig:polypolar}
\end{figure}

For \defm{polytropic} (\defm{calorically perfect}) equation of state (eos) the shock relations are easy to solve, with an abundance of explicit formulas. 
In particular, there is the following classical formula for the shock polar (see \cite{meyer-fan-ramp,busemann-handbuch-experimentalphysik}, or \cite[par.\ 121]{courant-friedrichs}):

\begin{alignat}{5} \uy = \pm |1-\ux| \sqrt{ \frac{\ux-\ux_n}{\ux_{\infty}-\ux} } \end{alignat}
where $\uu_0=(1,0)$ is scaled and rotated upstream velocity, $\uu=(\ux,\uy)$ downstream velocity, $1<\gisen<∞$ ratio of heats, $\Mach_0$ upstream Mach number, 
\begin{alignat}{5} \ux_n = \frac{\gisen-1+2\Mach_0^{-2}}{\gisen+1} \end{alignat}
the $\ux$ for a normal shock, 
\begin{alignat}{5} \ux_{\infty} = 1 + \frac{2\Mach_0^{-2}}{\gisen+1} \end{alignat}
the limit $\ux$ of the (unphysical) expansion branch.

From such formulas it is easily seen that the compressive part of the shock polar is a convex curve in the $\uu$ plane, 
so that it has ``standard'' behaviour:\\
1. each half has a unique critical shock, and \\
2. the critical and strong shocks are transonic. \\
The weak ones near critical are also transonic, but the range is usually narrow, separated by a \defm{sonic point} from the much larger segment of supersonic weak shocks.

The polytropic case is too restrictive since, to give one example, $\cvspec$ for oxygen cannot be treated as constant beyond a few hundred Kelvin above room temperature (\cite[fig.\ 5]{elling-idealpolar}). 
Of course many applications easily exceed this temperature range; for example ongoing research on hypersonic engines and/or combustion \cite{fang-zhang-deng-teng-acetylene-pof2019,li-chang-xu-yu-bao-song-pof2017,wang-zhang-yang-teng-pof2020}, or most obviously astrophysics \cite{linial-sari-pof2019}.

Shock polars for general eos have been discussed by many other authors, for example \cite{teshukov-polar} or \cite{henderson-menikoff}.
There the focus is generally on positive results rather than counterexamples or specifics of van der Waals eos. 

There is a much wider range of temperatures in which the eos is not polytropic but nevertheless ideal (thermally but not calorically perfect):
\begin{alignat}{5} \frac{\pp\idens}{\temp}=\Rspec=\text{constant},  \label{eq:idealgas} \end{alignat}
with specific gas constant $\Rspec$, volume per mass $\idens$, temperature $\temp$, linked to energy per mass by
\begin{alignat}{5} \epm=\epmt(\temp), \end{alignat}
where $\epmt$ is now a general function, in contrast to the classical calorically perfect case $\epmt(\temp)=\cvspec\temp$ with constant $\cvspec$. 
This case is significantly more difficult, because the wealth of explicit formulas is replaced by a few semi-explicit ones and a lot of implicit reasoning.
Nevertheless \cite{elling-idealpolar} recently found that ideal shock polars are ``standard'' in the sense above if the eos is convex, i.e.\ if $\pp$ a convex function of $\idens$ at constant entropy per mass $\spm$, assuming a few other reasonable conditions (e.g.\ specific heat $\cvspec>0$). 
This is satisfactory because any non-convex eos permits multiple compressive \emph{normal} shocks, so that the polar cannot be standard.

Although it is believed that there are physical fluids with eos non-convex in some regions of phase space \cite{lambrakis-thompson-1972,chandrasekharan-mercier-colonna-pof2021},
it is not clear whether the question is considered settled.
In any case, since most gases have convex eos, the results permit using freely that ideal polars are ``standard'', in particular that small turning angles $\turn$ always have a unique weak reflection, eliminating the need to perform numerical computations for each concrete eos and each upstream state.

This is rather convenient since the preconditions of the result are simple. It is natural to wonder whether this convenience extends to non-ideal eos.
Although counterexamples can be given with \emph{artificially} constructed non-ideal eos (see \cite[sect.\ 10]{elling-idealpolar}), physically relevant eos remain to be discussed.

In the thermodynamic phase plane the region of ideal gas behaviour has as its high-temperature boundary the onset of dissociation and ionization.
(Some authors do not consider dissociation to be a non-ideal effect, as chemically reacting mixtures are still ideal in the modified sense
\begin{alignat}{5} \pp=\Runiv\temp n \end{alignat}
for $\Runiv$ universal gas constant, $n$ particle density in moles; since mass per particle is usually not constant during reactions, neither is $\Rspec$ in \eqref{eq:idealgas}).
\cite{elling-idealpolar} performed some preliminary numerical studies of dissociating diatomic gas which were inconclusive.

In contrast, the \emph{low}-temperature boundary of the ideal behaviour region is usually at the onset of significant inter-particle forces, at moderate pressures near transitions from gas to condensed states. This boundary is more complicated, especially near the thermodynamic critical point (not to be confused with the critical points of shock polars). Some of the most important fluids, such as water or carbon dioxide (\cite[fig.\ 1]{wang-hickey-pof2020}), are particularly difficult to model by reasonably simple eos.
As is standard in the literature, we choose the van der Waals eos as a representative of non-ideal eos. Since it has few parameters, any pathologies possible despite this lack of freedom are likely to be present in many other non-ideal eos with more generous parameter space as well. Indeed we find that ``standard'' behaviour does not hold even for this simplest model, which brings the discussion of the lower temperature boundary to a conclusion.

\section{Van der Waals specifics}

The ``incomplete'' van der Waals eos is
\begin{alignat}{5} \rtemp = \big(\pp+\frac{\awaals}{\idens^2}\big)(\idens-\bidens) , \end{alignat}
in pressure form
\begin{alignat}{5} \pp = \frac{\Rspec\temp}{\idens-\bidens} - \frac{\awaals}{\idens²}  ; \end{alignat}
we abbreviate $\rtemp=\Rspec\temp$ and $\spms=\spm/\Rspec$ for specific gas constant $\Rspec$. 
For rarefied gas ($\idens\conv\infty$) the behaviour is ideal ($\rtemp\approx\pp\idens$). 
Using $\rtemp=\dats\epm\spms\idens$ and $\pp=-\dats\epm\idens\spms$ the method of characteristics yields a ``complete'' eos
\begin{alignat}{5} \epm = \epmx\big(\spms-\ln(\idens-\bidens)\big) - \frac{\awaals}{\idens}  \label{eq:complete-sv} \end{alignat}
for some function $\epmx$.
A particularly important special case is $\gisen$-van der Waals:
\begin{alignat}{5} \epmx(x) = \exp((\gisen-1)x) \end{alignat}
with $\gisen>1$; this corresponds to a van der Waals gas that is not only ideal but polytropic when rarefied.

We only consider positive temperature states:
\begin{alignat}{5} \rtemp = \datl\epm\spms\idens = \epmx'(\spms-\ln(\idens-\bidens)) > 0 . \end{alignat}

Thermodynamic stability requires that $\epm''$ is a positive definite matrix, i.e.
\begin{alignat}{5} 0 &< \ddatl\epm\spms\idens = \epmx''  \quad\text{and}\\
0 &< \Det\frac{\partial^2\epm}{\partial(\idens,\spms)^2}
\\&= (\frac{\epmx''+\epmx'}{(\idens-\bidens)^2}-2\frac{\awaals}{\idens^3}) \epmx'' - (\frac{\epmx''}{\idens-\bidens})^2
\\&= \big(\subeq{\epmx'}{=\rtemp}/(\idens-\bidens)^2-2\frac{\awaals}{\idens^3}\big) \subeq{\epmx''}{>0} . \end{alignat}
The first condition is equivalent to heat capacity $c_v$ at constant volume being positive:
\begin{alignat}{5} \datl\epm\rtemp\idens = \frac{\dats\epm\spms\idens}{\dats\rtemp\spms\idens}\ =\ \frac{\rtemp}{\epmx''} > 0 . \end{alignat}
By positive temperature this is essentially equivalent to $\epmx''>0$. This is satisfied in the $\gisen>1$ case, but needs to be required in the general non-$\gisen$ case.
The second condition then amounts to
\begin{alignat}{5} \rtemp &> 2\awaals \frac{(\idens-\bidens)^2}{\idens^3} . \end{alignat}
The curve of equality is called \defm{spinodal} (fig.\ \ref{fig:supervt}).

$\epmx''>0$ means $\epmx'$ is an invertible function, so we have
\[ \epmx(\spms-\ln(\idens-\bidens))=\epmt(\temp) \]
for a corresponding function $\epmt$. Under the assumptions made, the complete eos \eqref{eq:complete-sv} takes the alternative form
\begin{alignat}{5} \epm = \epmt(\temp) - \frac{\awaals}{\idens} . \end{alignat}
Here $\epmt$ describes the ideal behaviour in the rarefied $\idens\conv\infty$ limit. 

``Convex eos'' refers to an eos for pressure as function of volume, with entropy held fixed:
\begin{alignat}{5} \pp = -\datl\epm\idens\spm = \frac{\epmx'}{\idens-\bidens} - \frac{\awaals}{\idens^2} \end{alignat}
satisfies the condition
\begin{alignat}{5} 0 < \ddatl\pp\idens\spm = \frac{\epmx'''+3\epmx''+2\epmx'}{(\idens-\bidens)^3} - 6 \frac{\awaals}{\idens^4} . \end{alignat}
This is a lower bound on $\epmx'''$.

The isotherms in the $\idens$-$\pp$ plane have slope
\begin{alignat}{5}
  \datl\pp\idens\temp
  =
  - \frac{\rtemp}{(\idens-\bidens)^2} + 2 \frac{\awaals}{\idens^3} .
\end{alignat}
This expression is zero on the spinodal curve.
As the expression is decreasing in $\rtemp$ for fixed $\idens$, it is negative at all $\idens>\bidens$ for $\rtemp$ above some value $\rtempc$, the critical temperature.
At the critical temperature the expression is maximal with value zero, which is therefore a joint zero with its derivative
\begin{alignat}{5}
  \ddatl\pp\idens\temp
  =
  2 \frac{\rtemp}{(\idens-\bidens)^3} - 6 \frac{\awaals}{\idens^4} .
\end{alignat}
Combining the two equations to eliminate $\rtemp$, a joint zero $\idens$ is found, giving critical volume, temperature and pressure
\begin{alignat}{5} \idensc = 3\bidens \csep \rtempc = \frac{8\awaals}{27\bidens} \csep \ppc = \frac{\awaals}{27\bidens^2} . \end{alignat}
Below the critical temperature, for a fixed temperature $\temp$ some pressures $\pp$ can be realized by three different $\idens$, which is clearly an unstable situation, 
so the van der Waals pressure $\pp$ cannot be considered accurate without modification.

The standard modification is the Maxwell equal-area rule, replacing part of the isotherm by a horizontal line from $(\idens_l,\overline\pp)$ to $(\idens_g,\overline\pp)$ with $\idens_g(\temp)>\idens_l(\temp)$. This amounts to constant pressure $\overline\pp(\temp)$ for $\idens$ between $\idens_l,\idens_g$,
  with both endpoints on the original isotherm. 
Those states $\idens$ form the \defm{coexistence region} in the $(\idens,\temp)$ plane, bounded by the \defm{binodal curve} (fig.\ \ref{fig:supervt}).
States in that region represent vapor-liquid mixtures in evaporation-condensation equilibrium.
The equal-area rule is 
\begin{alignat}{5} \int_{\idens_l}^{\idens_g} \pp(\idens,\temp) d\idens = (\idens_g(\temp)-\idens_l(\temp))\overline\pp(\temp) ; \end{alignat}
where the left-hand side uses the unmodified van der Waals $\pp$. The rule can be derived from equilibrium requiring equal chemical potentials.

Our computations do not use the equal-area rule; instead we try to avoid the coexistence region altogether.

\section{Shocks}
\label{section:generaltools}%

Normal steady shocks are determined as solutions of the \defm{Hugoniot relation}
\begin{alignat}{5} \jump{\epm}+\avg{\pp}\jump{\idens} = 0 , \end{alignat}
where $\jump{f}=f-f_0$ is jump from upstream value $f_0$ to downstream value $f$, with average $\avg f=(f+f_0)/2$. 
For each fixed upstream thermodynamic state $(\tempu,\idensu)$, the \defm{Hugoniot curve} is the curve of downstream states $(\temp,\idens)$ solving the Hugoniot relation.
Given the two thermodynamic states, every other physical variabla for normal steady shocks can be derived, starting with
\begin{alignat}{5} \jnq = \frac{\jump\pp}{\jump{-\idens}} \end{alignat}
for normal mass flux $\jn=\jnu>0$, then normal velocities 
\begin{alignat}{5} \un = \idens\jn \csep \unu=\idensu\jnu, \end{alignat}
etc.
These equations are equivalent to conservation of mass, normal momentum and energy,
\begin{alignat}{5} 0 &= \jump{\dens\un} ,
\\ 0 &= \jump{\dens\unq+\pp},
\\ 0 &= \jump{\dens\un(\frac{\unq}{2}+\epm)+\un\pp} , \end{alignat}
where $\dens=1/\idens$ is mass density. 

For oblique shocks, with tangential velocity $\ut$, conservation of tangential momentum
\begin{alignat}{5} 0=\jump{\dens\un\ut} \end{alignat}
additionally yields  $\ut=\utu $.
The shock polar is the curve of oblique shocks with fixed $\uuu$ in addition to fixed $\tempu,\densu$.
We may simply obtain it from the Hugoniot curve by adding 
\begin{alignat}{5} \ut = \utu = \pm\sqrt{|\uuu|²-\unq}  \end{alignat}
(choice of $+$ or $-$ selects one half of the shock polar). From $\ut$ every other oblique-shock quantity can be computed; most important for our purposes is the 
deflection angle
\begin{alignat}{5} \turn=\betu-\bet \end{alignat}
where 
\begin{alignat}{5} \bet = \arcsin\frac{\un}{|\uu|} \end{alignat}
is angle between shock and downstream velocity $\uu$,
$\betu = \arcsin(\unu/|\uuu|) $ the same upstream. 

A point on the polar is called \defm{critical} if $\turn$ has a local extremum there.
In between extrema, in direction of increasing $\unu$ (i.e.\ increasing shock strength), 
segments running towards a local maximum of $|\turn|$ are called \defm{weak-type},
otherwise \defm{strong-type}. 
These definitions are necessary since we discuss polars with multiple critical points (fig.\ \ref{fig:closeup53}), so that some deflection angles $\turn$ allow three or more reflected shocks,
including multiple weak-type shocks, some of which can be stronger than some of the strong-type shocks. 

We define $\dhugs{f}{g}$ as the derivative of $f$ with respect to $g$ \emph{along the polar}. For pure normal-shock quantities this is the same as ``along the Hugoniot curve''.
Note that in contrast to derivatives $\dats{f}{g}{h}$ for some functions $f,g,h$ of the local state $(\temp,\idens)$,
here the derivative also depends on the chosen upstream state $\tempu,\idensu$. 

At critical points, where $\turn$ has an extremum, 
\begin{alignat}{5} \dhugl{\turn}{\betu} = 0. \end{alignat}
To find critical points fast numerically we use the equivalent formula 
\begin{alignat}{5} \utq \overset{\text{critical}}{=} \frac{\un(\unu-\un)}{1-\dhugs{\un}{\unu}} . \label{eq:utcrit} \end{alignat}
The formula is purely geometric, with all thermodynamics contained in the value of $∂\un/∂\unu$, which can be obtained as a formula involving the eos.
Put differently, given a point on the Hugoniot curve (normal shock), to find a polar that has upstream state $(\temp_0,\idens_0)$ and a critical point with downstream state $(\temp,\idens)$, choose
\begin{alignat}{5} |\uuu|² = \utq + \unuq = \frac{\un(\unu-\un)}{1-\dhugs{\un}{\unu}} + \unuq . \label{eq:q0crit} \end{alignat}
Importantly, this is the \emph{unique} $|\uuu|$ permitting a polar with such a critical point. This reduces the search space considerably.

\section{Restrictions}
\label{section:restrictions}

Using formula \eqref{eq:utcrit}, for simple non-ideal eos, for example $\gisen$-van der Waals, a numerical search can be performed by iterating over $\gisen$, upstream states $\temp_0,\idens_0$, and one Hugoniot curve parameter,
for example $\idens$. It is then easy to produce a large variety of critical points of polars with all kinds of pathologies. However, such cases are not interesting unless they are outside undesirable regions of phase space where the model or its shocks are unphysical or unstable, or where normal shocks already have similar pathologies. Compelling examples should satisfy several conditions.

  First, we do not permit any encounter with the region of non-convex eos, i.e.\ thermodynamic states $(\temp,\idens)$ where
  \begin{alignat}{5} \ddatl{\pp}{\idens}{\spm} ≤ 0. \end{alignat}
  Non-convex eos generally already permit multiple \emph{normal} steady shocks, as well as downstream \emph{normal} Mach numbers above $1$,
  so that discussion of oblique shocks is secondary. Besides, many of the shocks need not be physically correct; 
  instead of a single attached oblique shock at (say) a solid wedge, the physical flow may well be a combination of shocks and expansion/compression fans.

  Second, we also require thermodynamic stability by not permitting the polars to cross the spinodal curve into the region where the $2\times 2$ matrix of second derivatives of $\epm=\epm(\idens,\spm)$ is no longer positive definite. 
  The spinodal curve contains the region of non-convex eos in its interior if $\gisen$ is large, for example $\gisen=7/5$, but not for $\gisen$ closer to $1$ (see fig.\ \ref{fig:supervt} for the $\gisen=1.05$ case). 

  Beyond stability, we try to find examples avoiding the coexistence region where multiple phases, such as vapor and liquid, are present in equilibrium.
  This requirement is debatable; in many settings condensation is much slower than flow speed, especially near the boundary of the coexistence region.
  But the relative speed of condensation and flow depends too much on the specifics of the setting, for example on the spatial scale of the flow. In any case we wish to avoid the complications and qualifications required for discussing speed of condensation.
  Since we discount results involving the coexistence region, numerical computations were accelerated by \emph{not} applying the equal-area rule or other corrections, which should be kept in mind in diagrams
  such as fig.\ \ref{fig:supervt}.
  
Finally, the admissibility conditions of Lax and Liu (\cite{liu-admissibility-memoir,zhao-mentrelli-ruggeri-sugiyama}) are enforced along the entire Hugoniot curve.

\section{Supersonic critical reflections in $\gisen$-van der Waals}
\label{section:supercrit}

\begin{figure}
  \centerline{\tiny\input{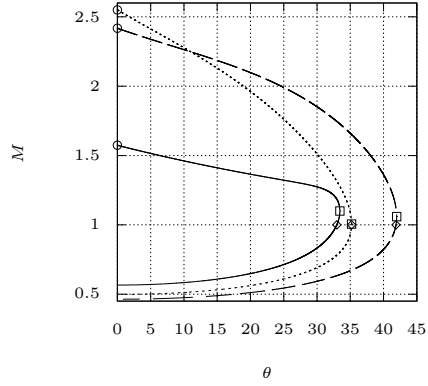}}
  \caption{Polars for $\gisen=1.05$ (solid curve), $\gamma=1.15$ (dashed), $\gamma=4/3$ (dotted); $\square$ indicates critical point, $\diamond$ sonic point}
  \label{fig:superMtheta}
\end{figure}

\begin{figure}
  \noindent\parbox{\linewidth}{%
    \tiny\input{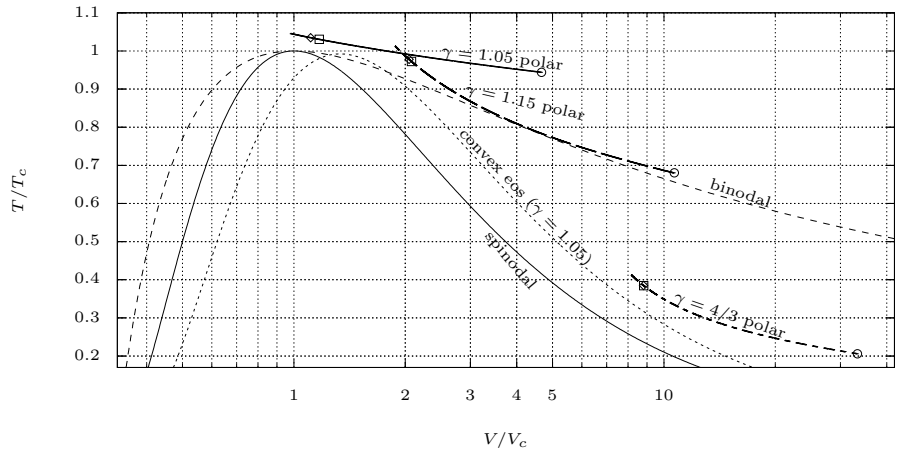}\caption{Hugoniot curves (bold) and coexistence, nonconvex and spinodal regions.}\label{fig:supervt}%
  }
\end{figure}

Despite the phase plane restrictions, polars with supersonic critical shocks can still be found. Figures \ref{fig:superMtheta}--\ref{fig:superbetturn} show examples for various $\gisen$ (see table \ref{fig:superparms} for parameters), with solid strokes representing $\gisen=1.05$, dashed for $\gisen=1.15$, dotted for $\gisen=4/3$; $\square$ indicates the critical point, $\diamond$ sonic point; $\circ$ is the vanishing shock ($\uu\conv\uuu$), whereas the unmarked end of each Hugoniot curve/polar is the normal shock. For large $\gisen$, namely $\gisen=4/3$, the Hugoniot curves are mostly located in the coexistence region (fig.\ \ref{fig:supervt}), and well below the critical temperature, but for $\gisen$ around $1.15$ and lower, examples outside the coexistence region are possible.

\begin{table}
\begin{tabular}{|l|l|l|l|l|l|l|l|}\hline
  line & $\gisen$ & $\idensu/\bidens$ & $\tempu/\tempc$ & $\Machu$ & $\Mach$ crit.\ & $\turn$ crit.\ & $\turn$ sonic \\\hline
  solid & 1.05 & $14$ & $0.9440$ & $1.5741$ & $1.1004$ & $33.4338^\circ$ & $32.9691^\circ$\\\hline
  dashed & 1.15 & $32$ & $0.6799$ & $2.4175$ & $1.0600$ & $41.9649^\circ$ & $41.8647^\circ$ \\\hline
  dotted & $4/3$ & $100$ & $0.2058$ & $2.5503$ & $1.006$ & $35.1956^\circ$ & $35.1942^\circ$ \\\hline
\end{tabular}
\caption{Parameters for the three supersonic polars in fig.\ \ref{fig:superMtheta}--\ref{fig:supervt}.}
\label{fig:superparms}
\end{table}

Fig.\ \ref{fig:superMtheta} shows the Mach numbers along each polar, plotted over turning angle $\turn$. All critical shocks (marked $\square$) are above the $\Mach=1$ grid line, i.e.\ in the supersonic region. This is also apparent in fig.\ \ref{fig:superuxuy} where the critical shocks are located right of the sonic shocks (marked $\diamond$) on each polar, whereas on ``standard'' shock polars they are located to the left. While critical and sonic point almost coincide in the $\gisen=4/3$ case, the separation is quite clear in the cases $\gisen=1.05$ and $\gisen=1.15$. Smaller $\gisen$ are typical for gas with higher temperature or more complex molecules, the latter having a wider range of non-ideal behaviour.

\begin{figure}
  \noindent\parbox{.47\linewidth}{%
    \centerline{\tiny\input{p-theta.xxx}}\caption{$\pp_0/\pp_c≈0.443$ for $\gisen=1.05$; $\pp_0/\pp_c≈0.0139$ for $\gisen=4/3$; $\pp_c=a/27b²$}\label{fig:superptheta}
  }%
  \hfil%
  \parbox{.47\linewidth}{%
    \centerline{\tiny\input{beta0-theta.xxx}}\caption{Angle $\betu$ between upstream velocity and shock}\label{fig:superbetturn}%
  }
\end{figure}

\begin{figure}
  \noindent\parbox{.47\linewidth}{\centerline{\tiny\input{ccot-g1.4.xxx}}
    \caption{$\csnd^2/\temp$ level sets from $4$ (leftmost dotted contour) to $1/4$ (lowest dotted) in factor $\sqrt{2}$ steps, for $\gisen=1.4$;
      solid contour spinodal, dashed binodal}
    \label{fig:ccot14}}%
  \hfil%
  \parbox{.47\linewidth}{\centerline{\tiny\input{ccot-g1.1.xxx}}
    \caption{Same as above, but for $\gisen=1.1$; sound speed increases rapidly as $\idens\dnconv\bidens$ (left boundary), decreases near the right side of the critical point}
    \label{fig:ccot11}}
\end{figure}

The examples are mostly located near the critical point or on the right side of the spinodal curves (cf.\ fig.\ \ref{fig:supervt}).
For this a natural explanation can be given. 
The Mach number
\begin{alignat}{5} \Mach = |\uu| / \csnd \end{alignat}
is a ratio of quantities that have different sensitivity to the eos. The numerator
\begin{alignat}{5} |\uu|=\sqrt{\utq+\unq}=\sqrt{|\uuu|^2-\unuq+\unq} \end{alignat}
with $\un=\jn\idens$ and $\unu=\jn\idensu$ and $\jn=\jump\pp/\jump{-\idens}$ is via
\begin{alignat}{5} \pp=-\datl\epm\idens\spm \end{alignat}
a function of \emph{first} derivatives of the equation of state $\epm=\epm(\idens,\spm)$. In contrast, 
\begin{alignat}{5} \dens^2\csnd^2 = \ddatl\epm\idens\spm \end{alignat}
so that the denominator $\csnd^2$ depends on a \emph{second} derivative. Hence $\csnd^2$ is one order more sensitive to rapid changes and near-singular behaviour of the eos.

The eos changes rapidly near the $\idens=\bidens$ boundary (left side of diagrams in fig.\ \ref{fig:ccot14} and \ref{fig:ccot11}),
but towards there $\csnd$ \emph{in}creases, so that $\Mach$ tends to decrease, apparently prohibiting values above $1$.
In contrast, $\csnd$ tends to decrease towards the spinodal curve, favoring change to $\Mach>1$.

Note that the $\uu$ plane polars are still convex (fig.\ \ref{fig:superuxuy}); likewise the $\pp$-$\turn$ plane polars in fig. \ref{fig:superptheta} have a ``standard'' shape. Fig.\ \ref{fig:superbetturn} shows the $\betu$ used to create each turning angle $\turn$.
\begin{figure}
  \centerline{\tiny\input{vx-vy.xxx}}\caption{$\uu$-plane polars, all convex}\label{fig:superuxuy}%
\end{figure}

Some of our Hugoniot curves (fig.\ \ref{fig:supervt}) are close to those of \cite[fig.\ 2]{bates-montgomery-physrevlett2000} (see also \cite{bates-montgomery-pof1999,bates-instability-pof2007}). There the interest is in shocks with multi-dimensional \emph{instability}.

\section{Multiple critical reflections in non-polytropic van der Waals}
\label{section:nuq}

Compared to finding van der Waals polars with supersonic but unique critical points, it is relatively harder to find examples with \emph{multiple} critical points. 
In fact for $\gisen$-van der Waals a detailed numerical search did not reveal any candidates.
The reason is not obvious, but by using a few parameters $\gisen,\idensu,\tempu$, as opposed to a function $\epmx$ that can be thought of as an infinity of parameters,
we may simply have restricted our search space too much to discover more pathologies. 

To find examples with multiple critical points it is necessary to use non-$\gisen$ functions $\epmx$.
Since there is an infinity of choices, some guesswork and experimentation are required.

We use the following idea: given a normal shock, \eqref{eq:q0crit} gives the unique value $q_0^c$ of $|\uuu|=\sqrt{\unuq+\utq}$ turning it into an oblique shock that is a critical point on a fixed-$\uuu$ polar. This critical $q_0^c$ is a function of the normal shock quantities:
\begin{alignat}{5} q_0^c = \frac{\un(\unu-\un)}{1-\dhugs{\un}{\unu}} + \unuq , \end{alignat}
where $\dhugs{\un}{\unu}$ is obtained by the implicit function theorem from the Hugoniot relation $\jump\epm=-\avg\pp\jump\idens$; 
it involves second derivatives of $\epmx$ through the sound speed $\csnd$ and Gr\"uneisen coefficient $\Grun$. 
Its derivative $\dhugs{q_0^c}{\betu}$,
can be obtained as a formula in normal shock quantities as well, which we do not give here as it is lengthy and its details unimportant,
other than being a linear function of the third derivative of $\epmx$. 
$\dhugs{q_0^c}{\betu}$ is usually positive, corresponding to increasing $q_0^c$ as the shock strength increases.
If, at any point, we can choose the eos-defining $\epmx$ to make $\dhugs{q_0^c}{\betu}$ negative,
then we return to smaller $q_0^c$ we have already visited, which means we have generated \emph{new} critical points on an \emph{old} polar.

The range of the third derivative $\epmx'''$ is constrained by our requirement that the eos is convex, i.e.\
\begin{alignat}{5} 0 < \fuder = \idens \frac{-(\partial^3\epm/\partial\idens^3)_\spm}{2\dens^2\csnd^2} . \end{alignat}
The numerator expands to 
\begin{alignat}{5} \frac{\epmx'''+3\epmx''+2\epmx'}{(\idens-\bidens)^2} - 6 \frac{\awaals}{\idens^4}, \end{alignat}
containing $\epmx'''$, which is therefore lower-bounded by $\fuder>0$.
Since $\dhugs{q_0^c}{\betu}$ is linear in $\epmx'''$, all we have to do is check numerically whether any $\epmx'''$ in the admissible interval makes $\dhugs{q_0^c}{\betu}<0$.
This numerical test is significantly faster than more straightforward methods. 

For our numerical search we chose to start the Hugoniot curve with $\epmx'''=(\gisen-1)\epmx''$, corresponding to a $\gisen$-van der Waals eos. 
The numerical test reveals quickly that for a wide range of $\gisen$ and upstream states $\idensu,\tempu$,
there are points along the Hugoniot curve where a different admissible $\epmx'''$ permits $\dhugs{q_0^c}{\betu}<0$. 

To fix a definite eos, we switch at some point to the unique $\epmx'''$ corresponding to constant $\fuder=0.01$, i.e.\ a slightly convex eos.
The switching point is arbitrary, but good results are obtained by switching as soon as $\dhugs{q_0^c}{\betu}<0$ is possible for \emph{some} admissible $\epmx'''$.
This point often occurs shortly after the critical shock on a $\gisen$-van der Waals polar. 

If $\fuder=0.01$ is sustained indefinitely, the polar will usually be incomplete, ending before a normal shock is reached,
for several possible reasons: $\epmx''<0$ so that heat capacity $\cvspec$ becomes negative, or spinodal curve reached, or admissibility criteria violated, etc.
Although there is no particular theoretical reason, some readers may regard complete polars as more convincing examples, so we chose to switch back to
$\epmx'''=(\gisen-1)\epmx''$, same as for the original $\gisen$-van der Waals, at some later point which was chosen arbitrarily, but early enough to reach the normal shock,
yet late enough to permit a second $|\turn|$ maximum.

Out of an infinity of cases we select one interesting example for display in fig.\ \ref{fig:nuq-theta53}. There we start with $\gisen=5/3$, $\idensu/\bidens=1.6$ and $\tempu/\tempc\approx0.851707$, with $\Machu\approx3.645829$;
the switch explained above is made at $\betu=52.797^\circ$ and the switch back at $\betu=66.0043^\circ$. The entire resulting Hugoniot curve (not shown) is outside the coexistence region. 
As desired we have found a polar with multiple critical points: $\turn$ has multiple maxima, which is easier to see in the close-up in fig.\ \ref{fig:closeup53}.
For $\turn=7.425^\circ$ there are four solutions, two weak-type ones ($\turn$ increasing as $\betu$ increases) and two strong-type ones. 
Naturally the second weak-type one has higher shock strength than the first strong-type one.

\begin{figure}
\parbox[t]{.47\linewidth}{%
  \tiny%
  \input{nuq53-beta0-theta.xxx}%
  \caption{Solid curve $\theta$ (see close-up in fig.\ \ref{fig:closeup53}); $\fuder=0.01$ between the vertical  (dotted-dashed vertical lines), $\gisen=5/3$ elsewhere. 
    $\idensu/\bidens=1.6$, $\tempu/\tempc\approx 0.851707$, $\Machu\approx 3.645829$. 
  }
  \label{fig:nuq-theta53}
}\hfil\parbox[t]{.47\linewidth}{%
  \tiny%
  \input{nuq53-beta0-theta-detail.xxx}%
  \caption{Detail: at $\theta=3.15^\circ$ there are four possible shocks, two weak-type (``w''), two strong-type (``s'').
  }
  \label{fig:closeup53}
}\\
\parbox[t]{.49\linewidth}{%
  \tiny%
  \input{nuq53-beta0-v.xxx}%
  \caption{Same as above: $v$ over $\beta_0$. 
  }
}\hfil%
\parbox[t]{.49\linewidth}{%
  \tiny%
  \input{nuq53-beta0-M.xxx}%
  \caption{Unlike the ideal gas case $M$ need not be monotone along the subsonic part of the polar}
  \label{fig:Mach53}
}\\
\parbox[t]{.49\linewidth}{%
  \tiny%
  \input{nuq53-beta0-Grueneisen.xxx}%
  \caption{Positive $\Grun$ throughout}
  \label{fig:Gruen53}
}\hfil%
\parbox[t]{.49\linewidth}{%
  \tiny%
  \input{nuq53-beta0-c.xxx}%
  \caption{$c$ not monotone}
}
\end{figure}

The $\turn$ gap between maxima and minimum that our choices achieve are quite small. This is in part because the eos reaches forbidden regions quickly if we sustain $\fuder=0.01$, 
in part because we modified a $\gisen$-van der Waals polar near its critical point, where $\turn$ differences are zero to first order in $\betu$ to begin with. It remains to be seen whether new ideas, or different non-ideal eos, can achieve a wider gap between $\turn$ extrema.

On the other hand the differences between (say) the two weak-type reflections are large in most other variables.
For example the shock-velocity angles $\betu$ differ by about $8^\circ$.

We also observe that the Mach number is no longer decreasing along the subsonic part of the polar, unlike the case of ideal convex eos \cite{elling-idealpolar}:
fig.\ \ref{fig:Mach53} shows a significant rise from about 0.75 to over 0.81 in the $\fuder=0.01$ region.

In this region the Gr\"uneisen parameter $\Grun$ drops from 3.5 to almost 0 (fig.\ \ref{fig:Gruen53}). To avoid reaching negative $\Grun$ the switch back to a $\gisen$-van der Waals eos $\epmx'''=(\gisen-1)\epmx''$ was performed.

The sound speed $\csnd$ is not monotone in this example. In fact we did not discover any examples with downstream sound speed $\csnd$ monotone along the entire Hugoniot curve. It is not clear whether such examples do not exist or whether some additional ideas are needed to construct them; after all the variety of possible functions $\epmx$ is infinite-dimensional.
On the other hand, it is possible to find examples of multiple critical points where $\fuder$ is not $0.01$ but some constant larger value, above $1$ or even $2$.
($\fuder>1$ does not force rising sound speed because it requires positive derivative of $\csnd$ along the \emph{isentrope},
not along the Hugoniot curve which deviates significantly for non-small shocks.)

\section{Consequences for theory}
\label{section:consequences}

\begin{figure}
  \centerline{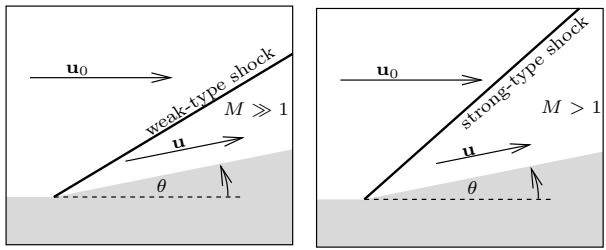}
  \caption{Two supersonic steady reflections for the same solid corner angle $\turn$, for values near the critical angle in fig.\ \ref{fig:superMtheta}}
  \label{fig:steady}
\end{figure}

If a critical point is supersonic, then for $\turn$ angles slightly below critical both strong- and weak-type reflections are supersonic.
In particular, in supersonic flow along a ramp both supersonic strong- and weak-type reflected shocks may be possible (fig.\ \ref{fig:steady}). 
This causes theoretical problems, since literature explanations for why the strong shock is unstable generally assume explicitly or implicitly that it is a \emph{transonic} shock, with subsonic downstream side.
Some explanations propose that any transonic shock is unstable because downstream perturbations can generate acoustic waves that reach the reflection point,
although \cite{teshukov} finds that weak-type transonic shocks are still stable (see \cite{elling-sonic-potf} for a similar conclusion but for structural rather than dynamic stability). 

The author believes that strong-type reflections will still be found unstable even when supersonic,
but a longer theoretical and experimental investigation will be necessary. 

Similarly, if some polar has multiple critical points, then some $\turn$ permit \emph{multiple weak-type} reflections, so that any arguments excluding strong-type shocks are insufficient to select a unique solution. Since theory gives some indication that weak-type shocks are stable (\cite{teshukov,elling-sonic-potf,elling-liu-rims05}), the author suspects that those reflections will be found stable under perturbations, but stability can be defined in multiple senses that may produce different answers; again a much longer discussion will be necessary. 

\begin{figure}
  \centerline{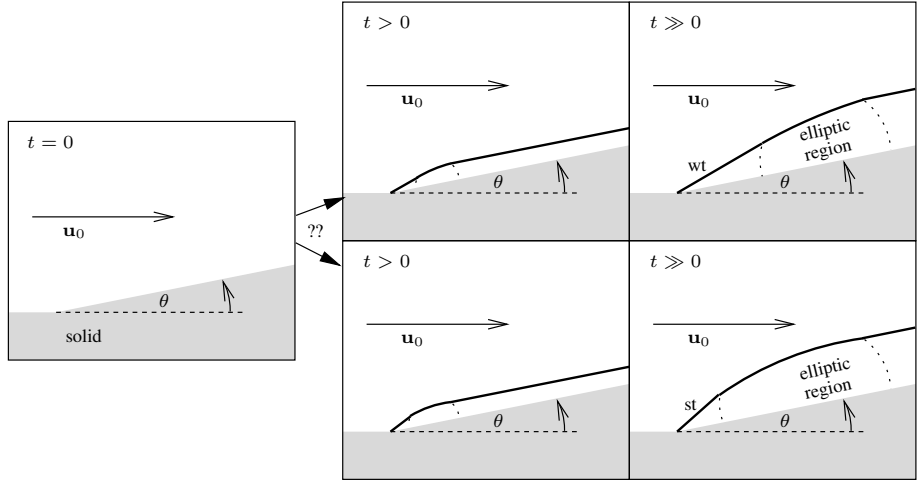}
  \caption{Self-similar reflections at a solid corner, for initial data $\vv_0,\dens_0$ constant in the entire domain.
    For $\turn$ near the critical shocks in fig.\ \ref{fig:superMtheta}, two different supersonic reflections (wt,st) \emph{may} be possible.}
  \label{fig:selfsim}
\end{figure}

Beyond steady reflection problems it is even more exciting to consider consequences for initial-value problems. \cite{elling-liu-pmeyer} considered supersonic potential flow onto a solid ramp,
with the entire region at initial time $t=0$ filled with upstream state (fig.\ \ref{fig:selfsim} left).
At $t>0$ a straight parallel shock separates from the ramp far from the corner;
at the corner some polar-determined reflected shock appears. The two are connected by a curved shock segment, with a region of non-constant flow below.  
The flow is self-similar, with $\vv$ and other flow variables functions of $(x/t,y/t)$ alone.
In the frame of an observer travelling at a fixed $(x/t,y/t)$ coordinate, the flow in the constant region is supersonic (hyperbolic), in the non-constant region it is subsonic (elliptic), with sonic (parabolic) segments connecting to the constant parts.

If there is more than one possible supersonic reflection, then there may be more than one self-similar wedge flow of this type. This would be an example of non-uniqueness for the initial-value problem; the initial data does not determine the future uniquely. However, the existence and construction of the non-constant regions is very nontrivial. \cite{elling-liu-pmeyer} solved the problem for compressible potential flow, a model that permits shocks but suppresses vorticity $\nabla\times\uu$; the model does not permit the pathologies uncovered in this paper. Construction of fig.\ \ref{fig:selfsim} and many other self-similar flow patterns for more complete models is still an open problem.

If there is more than one \emph{weak-type} reflection, say transonic as in fig.\ \ref{fig:closeup53},
then each of them may be included in a flow that is similar except for nonconstant elliptic regions extending all the way to the corner.
Which of these flows can actually occur is also unclear. 
For example, the curved shock needed to connect the corner shock to the ramp-parallel shock necessarily passes through a range of tangents,
which would necessarily include critical and strong-type ones if the upstream state was the same for all the shocks. 
But due to the self-similar nature of the flow the shock polars are different in each point (the upstream velocity is shifted by subtracting the $x/t,y/t$ coordinate). Although each corner shock is certainly possible as a local solution, it is unclear which of the curved connectors are possible in a global flow.

In case several reflections are possible, a next step is to consider additional stability criteria.
Our examples satisfy the weak stability criterion (\cite[sec.\ 15.2.1]{benzoni-gavage-serre}, see also \cite{majda-stability,dyakov-stability,kontorovich-stability,erpenbeck-stability,swan-fowles-pof1975,fowles-jfm1981}), and the example with multiple critical points also satisfies the uniform stability criterion. These conditions are essentially inviscid, so it is not clear, especially in the case of strong shocks, whether viscosity and heat conduction terms can restore or destroy stability. For compressible flow the corresponding parabolic coefficients are a bit too varied for a definite conclusion, to say nothing of boundary layers, or kinetic effects due to significant deviation from thermal equilibrium when shock strength is larger.

In other settings multiple solutions are really possible in a physical fluids, in particular in detonations or absence thereof \cite{radulescu-mevel-xiao-gallier}.
For instance in a fuel-oxidizer mixture some parameters allow multiple normal shocks with same upstream state, one shock causing detonation, the other staying below ignition temperature. Which of these choices occurs can obviously not be decided by considering shock relations in a purely inviscid model; the question of ignition is complex. It is nevertheless important in many applications, for example ramjets/scramjets
(\cite{fang-zhang-deng-teng-acetylene-pof2019,li-chang-xu-yu-bao-song-pof2017,wang-zhang-yang-teng-pof2020}).
Of course the analogy between van der Waals effects and combustion cannot be carried too far, since the latter is an irreversible process.

\section{Conclusion}
\label{section:conclusions}

We found that in contrast to the ideal eos case settled in \cite{elling-idealpolar},
both supersonic and multiple critical points are possible on shock polars for the most common non-ideal eos, the van der Waals model,
even when the entire polar avoids the undesirable regions from section \ref{section:restrictions}. 
To find examples with multiple critical points it is necessary to consider a general (non-$\gisen$) van der Waals eos.
The transition to and from constant fundamental derivative $\fuder=0.01$ is abrupt, so it is unclear whether some physical fluid has transitions sufficiently sharp
to allow multiple critical points.

In contrast, for supersonic critical points $\gisen$-van der Waals suffices;
$\gisen$ below 1.15 are sufficient to avoid coexistence and other forbidden regions,
although we caution that coexistence regions of some physical fluids can deviate significantly from the van der Waals model (cf.\ \cite[fig.\ 2]{guggenheim-1945}). 
But the range of examples found is large enough that we expect some physical fluid near its thermodynamic critical point
to allow supersonic strong-type/critical shocks.

Direct experimental confirmation seems to require producing shock waves with compression ratios of 3 to 5 (fig.\ \ref{fig:supervt}) in a high--pressure fluid,
which is not easy. However, it would already be interesting to obtain \emph{indirect} confirmation, without generating shock waves, by \\
1. obtaining a sufficiently large set of eos data for a homogeneous non-moving fluid, by experiment or by molecular simulation, \\
2. then calculating numerically that some polar for this eos has supersonic critical shocks. \\
Again hydrocarbons seem promising since their economic importance has ensured richer experimental data and more refined molecular simulation models. 

However, our more immediate conclusion is on the theoretical side:
non-ideal gas appears to offer little hope for a catch-all theorem like the one for ideal gas in \cite{elling-idealpolar},
namely that convex eos and a few standard assumptions guarantee subsonic and unique critical shocks, as well as decrease of Mach number on subsonic parts of polars. 

Finally, our results show a need to revisit the weak-strong reflection problem,
namely to discuss supersonic strong-type reflections, a possiblity that appears to have been ignored in the literature.

\section*{Acknowledgements}

This research was partially supported by Taiwan MOST Grant No.\ 110-2115-M-001-005-MY3.

\bibliographystyle{amsalpha}

\begin{thebibliography}{RMXG21}

\bibitem[Bat07]{bates-instability-pof2007}
{J.W.} Bates, \emph{Instability of isolated planar shock waves}, Phys. Fluids
  \textbf{19} (2007), no.~094102.

\bibitem[BD92]{ben-dor-book}
{G.} Ben-Dor, \emph{Shock wave reflection phenomena}, Springer, 1992.

\bibitem[BGS07]{benzoni-gavage-serre}
{S.} Benzoni-Gavage and {D.} Serre, \emph{Multi-dimensional hyperbolic partial
  differential equations}, Oxford University Press, 2007.

\bibitem[BM99]{bates-montgomery-pof1999}
{J.W.} Bates and {D.C.} Montgomery, \emph{Some numerical studies of exotic
  shock wave behaviour}, Phys. Fluids \textbf{11} (1999), 462--475.

\bibitem[BM00]{bates-montgomery-physrevlett2000}
\bysame, \emph{The {D'yakov}-{Kontorovich} instability of shock waves in real
  gases}, Phys. Rev. Letters \textbf{84} (2000), no.~6, 1180--1183.

\bibitem[Bus31]{busemann-handbuch-experimentalphysik}
{A.} Busemann, \emph{Handbuch der {Experimentalphysik}}, vol.~{IV}, Akademische
  Verlagsgesellschaft, Leipzig, 1931.

\bibitem[CF48]{courant-friedrichs}
{R.} Courant and {K.O.} Friedrichs, \emph{Supersonic flow and shock waves},
  Interscience Publishers, 1948.

\bibitem[CMC21]{chandrasekharan-mercier-colonna-pof2021}
{N.B.} Chandrasekharan, {B.} Mercier, and {P.} Colonna, \emph{Nonlinear wave
  propagation in dense vapor of {Bethe}-{Zel'dovich}-{Thompson} fluids
  subjected to temperature gradients}, Physics of Fluids \textbf{33} (2021),
  107109.

\bibitem[D'i54]{dyakov-stability}
{S.P.} D'iakov, Zh. Eksp. Teor. Fiz. \textbf{27} (1954), no.~3, 288--295.

\bibitem[EL06]{elling-liu-rims05}
{V.} Elling and Tai-Ping Liu, \emph{Physicality of weak {Prandtl}-{Meyer}
  reflection}, RIMS Kokyuroku, vol. 1495, Kyoto University, Research Institute
  for Mathematical Sciences, May 2006, pp.~112--117.

\bibitem[EL08]{elling-liu-pmeyer}
\bysame, \emph{Supersonic flow onto a solid wedge}, Comm. Pure Appl. Math.
  \textbf{61} (2008), no.~10, 1347--1448.

\bibitem[Ell09]{elling-sonic-potf}
V.~Elling, \emph{Counterexamples to the sonic criterion}, Arch. Rat. Mech.
  Anal. \textbf{194} (2009), no.~3, 987--1010.

\bibitem[Ell10]{elling-rrefl-lax}
{V.} Elling, \emph{Regular reflection in self-similar potential flow and the
  sonic criterion}, Commun. Math. Anal. \textbf{8} (2010), no.~2, 22--69.

\bibitem[Ell21]{elling-idealpolar}
\bysame, \emph{Shock polars for ideal and non-ideal gas}, J. Fluid Mech.
  \textbf{916} (2021), no.~A51.

\bibitem[Erp62]{erpenbeck-stability}
{J.J.} Erpenbeck, \emph{Stability of step shocks}, Phys. Fluids \textbf{5}
  (1962), no.~10, 1181--1187.

\bibitem[Fow81]{fowles-jfm1981}
{G.R.} Fowles, \emph{Stimulated and spontaneous emission of acoustic waves from
  shock fronts}, Phys. Fluids \textbf{24} (1981), 220--227.

\bibitem[FZDT19]{fang-zhang-deng-teng-acetylene-pof2019}
{Y.} Fang, {Y.} Zhang, {X.} Deng, and {H.} Teng, \emph{Structure of
  wedge-induced oblique detonation in acetylene-oxygen-argon mixtures}, Phys.
  Fluids \textbf{31} (2019), no.~026108.

\bibitem[Gug45]{guggenheim-1945}
{E.A.} Guggenheim, \emph{The principle of corresponding states}, J. Chem. Phys.
  \textbf{13} (1945), no.~7, 253--261.

\bibitem[HM98]{henderson-menikoff}
{L.F.} Henderson and {R.} Menikoff, \emph{Triple-shock entropy theorem and its
  consequences}, J. Fluid Mech. \textbf{366} (1998), 179--210.

\bibitem[Hor86]{hornung-reviews}
{H.} Hornung, \emph{Regular and {Mach} reflection of shock waves}, Ann. Rev.
  Fluid Mech. \textbf{18} (1986), 33--58.

\bibitem[Kon57]{kontorovich-stability}
{V.M.} Kontorovich, \emph{Concerning the stability of shock waves}, J. Exptl.
  Theoret. Phys. (U.S.S.R.) \textbf{33} (1957), 1525--1526.

\bibitem[LCX{\etalchar{+}}17]{li-chang-xu-yu-bao-song-pof2017}
{N.} Li, {J.-T.} Chang, {K.-J.} Xu, {D.-R.} Yu, {W.} Bao, and {Y.-P.} Song,
  \emph{Prediction dynamic model of shock train with complex background waves},
  Phys. Fluids \textbf{29} (2017), no.~116103.

\bibitem[Liu81]{liu-admissibility-memoir}
Tai-Ping Liu, \emph{Admissible solutions of hyperbolic conservation laws},
  Memoirs AMS, no. 240, American Mathematical Society, 1981.

\bibitem[LS19]{linial-sari-pof2019}
{I.} Linial and {R.} Sari, \emph{Oblique shock breakout from a uniform density
  medium}, Phys. Fluids \textbf{31} (2019), 097102.

\bibitem[LT72]{lambrakis-thompson-1972}
{K.C.} Lambrakis and {P.A.} Thompson, \emph{Existence of real fluids with a
  negative fundamental derivative {$\Gamma$}}, Phys. Fluids \textbf{15} (1972),
  no.~5, 933--935.

\bibitem[Maj83]{majda-stability}
{A.} Majda, \emph{The stability of multi-dimensional shock fronts}, vol. 273,
  AMS, 1983.

\bibitem[Mey08]{meyer-fan-ramp}
{Th.} Meyer, \emph{{Ueber} zweidimensionale {Bewegungsvorg\"ange} in einem
  {Gas}, das mit {Ueberschallgeschwindigkeit} {str\"omt}}, {Forschungsheft} des
  {Vereins} {Deutscher} {Ingenieure} {(VDI)} \textbf{62} (1908), 31--67.

\bibitem[RMXG21]{radulescu-mevel-xiao-gallier}
{M.I.} Radulescu, {R.} {M\'{e}vel}, {Q.} Xiao, and {S.} Gallier, \emph{On the
  self-similarity of diffracting gaseous detonations and the critical channel
  width problem}, Phys. Fluids \textbf{33} (2021), 066106.

\bibitem[SF75]{swan-fowles-pof1975}
{G.W.} Swan and {G.R.} Fowles, \emph{Shock wave stability}, Phys. Fluids
  \textbf{18} (1975), no.~1, 28--35.

\bibitem[Tes86]{teshukov-polar}
{V.M.} Teshukov, \emph{On the shock polars in a gas with general equations of
  state}, J. Appl. Math. Mech. \textbf{50} (1986), no.~1, 71--75.

\bibitem[Tes89]{teshukov}
\bysame, \emph{Stability of regular shock wave reflection}, Prikl. Mekhanika i
  Techn. Fizika \textbf{30} (1989), no.~2, 26--33, translated in Appl. Mech.
  Tech. Phys. 30 (189) 1989.

\bibitem[vN43]{neumann-1943}
{J.} von Neumann, \emph{Oblique reflection of shocks}, Tech. Report~12, Navy
  Dep., Bureau of Ordnance, Washington, D.C., 1943, In: Collected works, v.\ 6,
  p.\ 238--299.

\bibitem[WH20]{wang-hickey-pof2020}
{J.} Wang and {J.-P.} Hickey, \emph{Analytical solutions to shock and expansion
  waves for non-ideal equations of state}, Phys. Fluids \textbf{32} (2020),
  086105.

\bibitem[WZYT20]{wang-zhang-yang-teng-pof2020}
{K.} Wang, {Z.} Zhang, {P.} Yang, and {H.} Teng, \emph{Numerical study on
  reflection of an oblique detonation wave on an outward turning wall}, Phys.
  Fluids \textbf{32} (2020), no.~046101.

\bibitem[ZMRS11]{zhao-mentrelli-ruggeri-sugiyama}
{N.} Zhao, {A.} Mentrelli, {T.} Ruggeri, and {M.} Sugiyama, \emph{Admissible
  shock waves and shock-induced phase transitions in a van der waals fluid},
  Phys. Fluids \textbf{23} (2011), no.~086101.

\end{thebibliography}
\newcommand{\etalchar}[1]{$^{#1}$}
\providecommand{\bysame}{\leavevmode\hbox to3em{\hrulefill}\thinspace}
\providecommand{\MR}{\relax\ifhmode\unskip\space\fi MR }
\providecommand{\MRhref}[2]{%
  \href{http://www.ams.org/mathscinet-getitem?mr=#1}{#2}
}
\providecommand{\href}[2]{#2}

\end{document}